\documentclass[%
reprint,
superscriptaddress,
amsmath,amssymb,
aps,
prl,
floatfix,
]{revtex4-1}

\usepackage{color}
\usepackage{graphicx}
\usepackage{dcolumn}
\usepackage{bm}
\usepackage{hyperref}
\usepackage[normalem]{ulem}

\hypersetup{
    colorlinks=true,
    linkcolor=blue,
    citecolor=blue,
    filecolor=black,
    urlcolor=blue,
}

\providecommand{\e}[1]{\ensuremath{\times 10^{#1}}}
\newcommand{\mpl}{M_\mathrm{Pl}}

\newcommand{\pd}{\partial}

\def\bk{{\mathbf{k}}}
\def\f {{\phi}}

\def\calR{{\cal R}}

\def\calS{{\cal S}}
\def\calP{{\cal P}}

\def\Pab{{\mathcal{P}_{\alpha \beta}}}

\def\PR{{\mathcal P_\calR}}
\def\PS{{\mathcal P_\calS}}

\def\|{{ \, || \,}}
\def\riso{r_\mathrm{iso}}
\def\kpiv{k_\mathrm{piv}}

\begin{document}

\title{Simple predictions from multifield inflationary models}

\author{Richard Easther}
  \email{r.easther@auckland.ac.nz}
  \affiliation{Department of Physics, University of Auckland, Private Bag 92019,  Auckland, New Zealand}

\author{Jonathan Frazer}
  \email{j.frazer@ucl.ac.uk}
  \affiliation{Department of Physics and Astronomy, University College London, London WC1E 6BT, U.K.}
  \affiliation{Department of Theoretical Physics, University of the Basque Country UPV/EHU, 48040 Bilbao, Spain
}

\author{Hiranya V. Peiris}
  \email{h.peiris@ucl.ac.uk}
  \affiliation{Department of Physics and Astronomy, University College London, London WC1E 6BT, U.K.}

\author{Layne C. Price}
  \email{lpri691@aucklanduni.ac.nz}
  \affiliation{Department of Physics, University of Auckland, Private Bag 92019,  Auckland, New Zealand}

\date{\today}

\begin{abstract}
 
  We explore whether multifield inflationary models make unambiguous predictions for fundamental cosmological observables. Focusing on  $N$-quadratic inflation, we numerically evaluate the full perturbation equations for models with 2, 3, and $\mathcal{O}(100)$ fields, using several distinct methods for specifying the initial values of the background fields.  All scenarios are highly predictive, with the probability distribution functions of the cosmological observables becoming more sharply peaked as $N$ increases. For  $N=100$ fields, 95\% of our Monte Carlo samples fall in the ranges  $n_s \in (0.9455,0.9534)$; $\alpha \in (-9.741,-7.047)\e{-4}$; $r\in(0.1445,0.1449)$; and $\riso \in (0.02137,3.510)\e{-3}$ for the spectral index, running, tensor-to-scalar ratio, and isocurvature-to-adiabatic ratio, respectively.  The expected amplitude of isocurvature perturbations grows with $N$, raising the possibility that many-field models may be sensitive to post-inflationary physics and suggesting new avenues for testing these scenarios.

\end{abstract}

\maketitle


The study of inflation has been transformed by the advent of precision cosmology. In 2013 the {\em Planck} Collaboration~\cite{Ade:2013zuv, Ade:2013uln} announced a $5 \sigma$ detection of scale-dependence in the  primordial power spectrum,  $\calP(k)$.  Likewise, the non-Gaussian component of the initial perturbations is less than   $0.01$\%~\cite{Ade:2013ydc} and there are strong limits on isocurvature perturbations~\cite{Ade:2013uln}.  These results are entirely consistent with single-field slow roll inflation.

The key theoretical challenge for inflation is to show how a phase of accelerated expansion is realized in particle physics. However, single-field models are not always natural; \emph{e.g.}, string compactifications often result in hundreds of scalar fields \cite{Grana:2005jc,Douglas:2006es,Denef:2007pq,Denef:2008wq}.  Multifield models yield a wider range of possible $\calP(k)$ and higher-order correlators than simple single-field scenarios. Consequently, it is vital to determine not only  what is \emph{possible} in multifield models, but whether specific multifield models   yield \emph{generic} predictions that can  be tested against  data.  

Multifield models permit many distinct inflationary trajectories, and can thus be sensitive to the initial values of the background fields.  The relative likelihood for different initial conditions (ICs) in the overall phase-space of the inflationary dynamical system  must be encoded in the Bayesian prior for the model. Inflationary models are, to some extent, {\em ad hoc} hypotheses, so the IC priors  typically cannot be computed or reliably predicted \emph{a priori}. Recently it was pointed out that some multifield models make predictions for the inflationary observables that do not depend strongly on the specific IC prior \cite{Frazer:2013zoa,Kaiser:2013sna,Kallosh:2013hoa,Kallosh:2013maa,Kallosh:2013daa}, and this class of model unambiguously predicts the distributions of the inflationary observables. On the other hand, observational data could constrain the initial field configuration for  models with strong sensitivity to their initial conditions. 

In this \emph{Letter} we present the first \emph{generic} predictions for a multifield inflation model in the many-field limit.  By numerically evolving the perturbations, we find the probability density functions (PDFs) for the spectral index $n_{s}$, the tensor-to-scalar ratio $r$, the running $\alpha$, and the isocurvature-to-adiabatic ratio $\riso$ in $N$-quadratic inflation.  We give the first complete analysis of the many-field case \cite{Liddle:1998jc,Kanti:1999ie,Easther:2005zr,Dimopoulos:2005ac,Kim:2006ys,Kim:2007bc} by exploring inflation with $N=100$ fields. We consider three distinct IC priors to assess the sensitivity of the model's predictions to the initial conditions.
 
We see novel behavior in the many-field case, where trajectories in field space ``turn'' until the end of inflation, yielding an increased $\riso$ that may be relevant to reheating.  The PDFs for  $n_s$, $\alpha$, and $r$ become more sharply peaked at large $N$, implying that the many-field case is predictive.  We also obtain high-density samples in the low-$N$ limit \cite{Polarski:1992dq,Ellis:2013iea} with $N=\{2,3\}$. In this limit we also see sharply peaked PDFs and a nontrivial consistency relation in the $(n_{s},\alpha)$--plane, but with a greater dependence on the IC prior than with $N=100$ fields.
 
{\bf Method:} We assess the ``predictivity" of  inflationary models as follows. We draw ICs from a specified prior probability distribution and evolve the background equations of motion. We require the pivot scale $\kpiv=0.002 \, \mathrm{Mpc}^{-1}$  to leave the horizon $55$  $e$-folds before the end of inflation; if there are fewer $e$-folds, we exclude the IC and draw another. Otherwise, we solve the perturbation equations numerically and compute observables by evaluating the power spectra at the end of inflation. Iterating this process, we obtain the PDF for the inflationary observables given the choice of IC prior.

We consider $N$-quadratic inflation with canonical kinetic terms,  minimal coupling to Einstein gravity, and  potential
\begin{equation}
V=\frac{1}{2} m_\alpha^2 \phi_{\alpha}^2 \, , 
\label{eqn:Vquad}
\end{equation}
with an implied sum over repeated field indices. This model makes an excellent test case as it is both extremely simple and well-defined for any value of $N$. 

For $N=100$ fields, we follow Ref.~\cite{Easther:2005zr} and draw the mass values from the Mar\v{c}enko-Pastur distribution with  $\beta=0.5$.  We choose the largest mass ratio as $1/8.08$ and the other masses so that they are equally spaced in the cumulative probability distribution function.  We do not expect our results to depend strongly on this  choice, provided the masses are all of the same order of magnitude.  

\begin{figure}
  \includegraphics{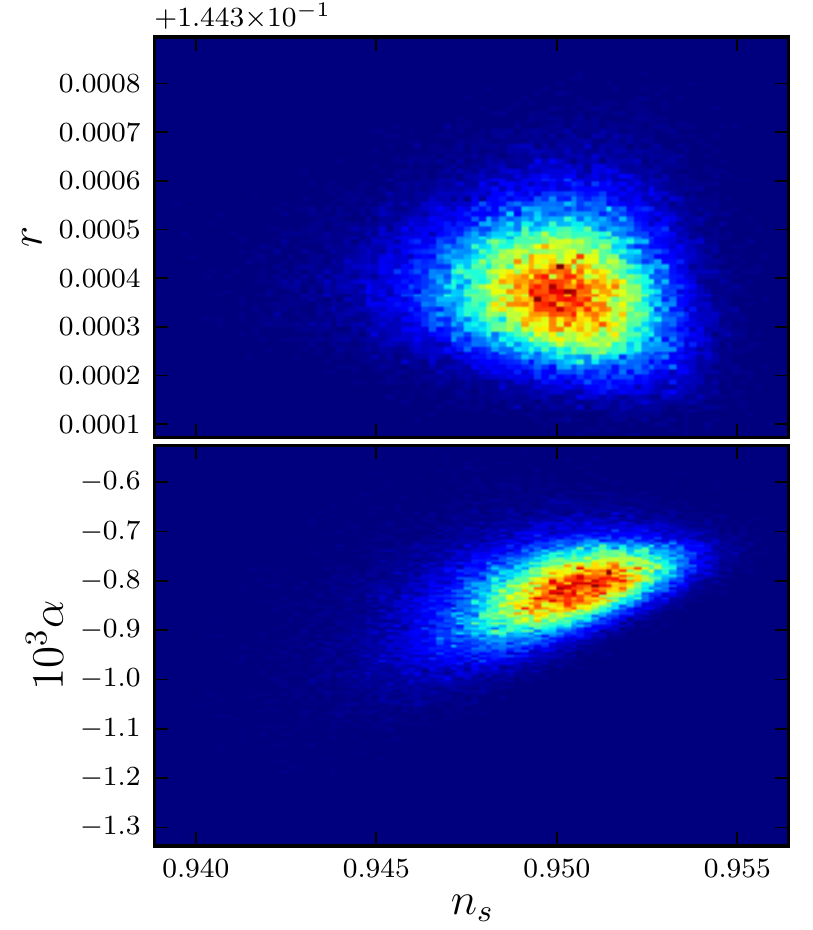}
  \caption{Histograms for $N=100$ fields with the iso-$E_0$ prior at $E_0=0.1\, \mpl$ and masses from the Mar\v{c}enko-Pastur distribution \cite{Easther:2005zr} with a maximum mass ratio of $1/8.08$ and $r$ is plotted relative to a baseline value of $0.1443$.  All observables are contained within a very small subvolume of the much larger range of possible values  the model can yield, showing that it makes precise predictions for the values of the inflationary observables.}
  \label{fig:nsralpha_100field_EE}
\end{figure}

\textbf{Initial conditions:}
We consider three IC priors:
\begin{enumerate}
\item The \emph{iso-$E_0$ prior} \cite{Easther:2013bga} with an equal-area prior on an initial surface with energy $E_0$.
\item The \emph{iso-$N_e$ prior} \cite{Frazer:2013zoa} with an equal-area prior set approximately $N_e$ $e$-folds before the end of inflation on the surface $\phi_\alpha \phi_\alpha = 4 N_e$.
\item The \emph{slow-roll prior} with velocities set in slow-roll and field ICs distributed uniformly over some pre-defined range.
\end{enumerate}

Each prior has a different physical justification and leads to significantly different distributions for the field values and velocities. For example, the iso-$N_e$ prior near $N_e=55$ implies we know nothing about the initial state when observable scales start to leave the horizon. By contrast, either (A) using the iso-$E_0$ prior with a relatively large initial energy; (B) requiring $N_e \gg 55$ for the iso-$N_e$ prior; or (C) specifying a large field-space range for the slow-roll prior  typically give solutions more scope to evolve into dynamically-favored regions of phase space, \emph{e.g.}, slow-roll along the direction of the lightest field. Consequently,  with these IC priors  a higher proportion of trajectories  find the attractors before the end of inflation.  Conversely, the iso-$N_e$ prior with $N_e \sim 55$ is the least predictive of these choices.

\textbf{Multifield perturbations:} We use an extended version of {\sc ModeCode}~\cite{Mortonson:2010er,Easther:2011yq,Norena:2012rs,MultiModeCode} that evolves the perturbation spectrum for an arbitrary potential, initial field values and velocities.  {\sc ModeCode} solves the 2-index mode equation~\cite{Salopek:1988qh,Huston:2011fr}
\begin{align}
  \label{eqn:mmeq}
  \psi_{\alpha\beta}^{\prime \prime} + (1-\epsilon) \psi_{\alpha\beta}^\prime +\left(\frac{k^2}{a^2H^2} - 2 + \epsilon \right) \psi_{\alpha\beta}
  \\
  + {\cal M}_{\alpha\gamma} \psi_{\gamma\beta} = 0 ~, \notag
\end{align}
where primes represent derivatives with respect to the number of $e$-folds, $N_{e}$; $\epsilon\equiv -\dot{H}/H^{2}$ is the slow-roll parameter; and $\psi_{\alpha\beta}$ is related to the Mukhanov--Sasaki variable, $u_{\alpha}\equiv a\delta\phi_{\alpha}$, by a sum over annihilation operators: $u_{\alpha}(\bk, N) = \psi_{\alpha\beta}(\bk, N) \hat a_{\beta}(\bk).$
Finally, the mass matrix is given by
\begin{equation}
  \mathcal M_{\alpha\beta} = \frac{\pd_{\alpha}\pd_{\beta} V}{H^2} + \frac{\left(\phi_{\alpha}^\prime \pd_{\beta} V  +  \phi_{\beta}^\prime \pd_{\alpha} V \right)
}{H^2}   + (3-\epsilon) \phi_{\alpha}^\prime \phi_{\beta}^\prime ,
  \label{eqn:cij}
\end{equation}
where the Hubble parameter is $ H^2 =  V/(3 - \epsilon)$.

For a mode $k$, we set the Bunch-Davies initial state for $\psi_{\alpha\beta}$ when $100 k = aH$.  The power spectrum for the field perturbations $\delta\phi_{\alpha}$ is 
\begin{equation}
  {\cal P}_{\alpha\beta} (k) = \frac{k^3}{2\pi^2} \left( \frac{1}{a^2} \right)\psi_{\alpha\gamma}(k)\psi^*_{\beta\gamma}(k) \, ,
  \label{eqn:powerspectrum}
\end{equation}
where star denotes complex conjugation. We compute the power spectra for the comoving curvature perturbation $\calR$ and isocurvature perturbations $\calS$ via an appropriately-scaled projection onto directions parallel and perpendicular to the background trajectory.  Consequently,
\begin{equation}
  \PR (k) = \frac{1}{2 \epsilon} \omega_\alpha  \omega_\beta {\cal P}_{\alpha\beta} (k) 
  \label{eqn:pad}
\end{equation}
where $\omega_\alpha= \phi_{\alpha}^\prime / \phi_0^\prime$ projects onto the direction of the background trajectory for $ \phi^{\prime \, 2}_{0} \equiv \f_{\alpha}^\prime \f_{\alpha}^\prime $. 
Directions perpendicular to $\omega$ are isocurvature directions, and can source superhorizon evolution of $\calR$;  we find the $(N-1)$ isocurvature vectors $s_I$ by Gram-Schmidt orthogonalization.  In analogy to $\calR$, we define isocurvature perturbations $\calS_I \equiv - (1/\phi_0^\prime) s_{I \alpha} \delta \phi_\alpha$  with the spectrum
\begin{equation}
  \PS (k) = \frac{1}{2 \epsilon}\sum_{I,J}^{N -1} s_{I \alpha} s_{J \beta} \Pab(k) \, .
  \label{eqn:piso}
\end{equation}

Conventionally, $\PR$ is characterized by an amplitude $A_s$ and its logarithmic derivatives $\mathcal D = d/d\log k$ at the pivot scale, with $n_s = \mathcal D \log \PR$ and $\alpha = \mathcal D^2 \log \PR$.  We can similarly describe $\PS$ or the adiabatic-isocurvature cross spectrum, although we report only the isocurvature-to-adiabatic ratio $\riso=\PS/\PR$. While \textsc{ModeCode} numerically computes the full functional form of $\PR(k)$ and $\PS(k)$, for convenience $n_s$ and $\alpha$ are computed by central finite differences near $\kpiv$.  Finally, we compute the tensor-to-scalar ratio $r$ by evolving the appropriately-normalized tensor perturbations.

\begin{figure}
  \includegraphics{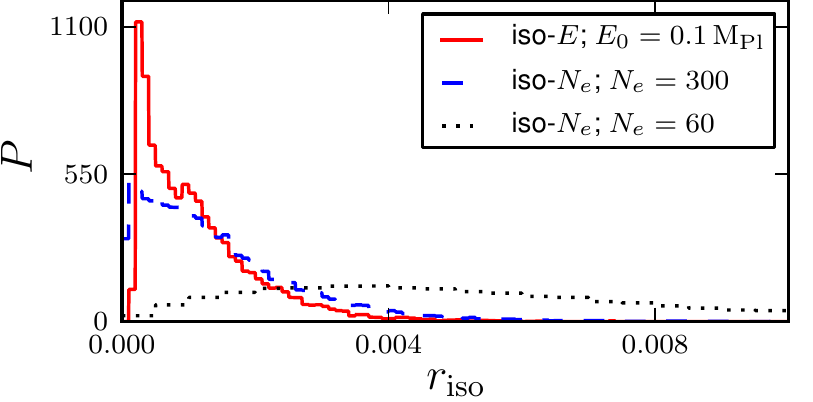}
  \caption{Density function of isocurvature fraction $\riso=\PS/\PR$ for different initial condition priors with $N=100$ fields.  The average $\riso$ roughly decreases with increasing number of $e$-folds between the surface on which we specify the IC and the end of inflation, implying the heavier masses find their minima more often when given more time to evolve before inflation ends.}
  \label{fig:r_iso}
\end{figure}

\begin{figure}
  \includegraphics[width=0.48\textwidth]{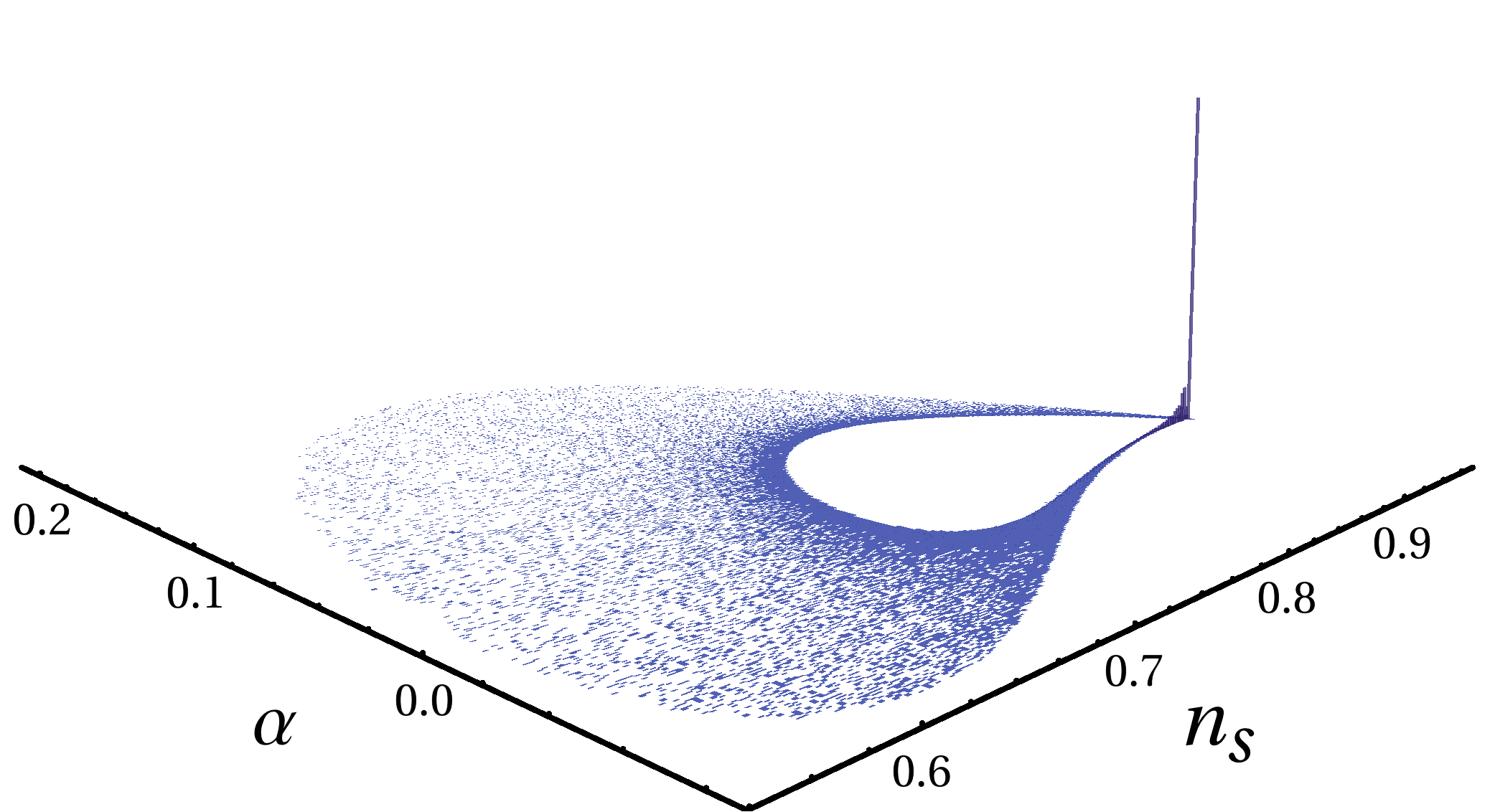}
  \includegraphics[width=0.48\textwidth]{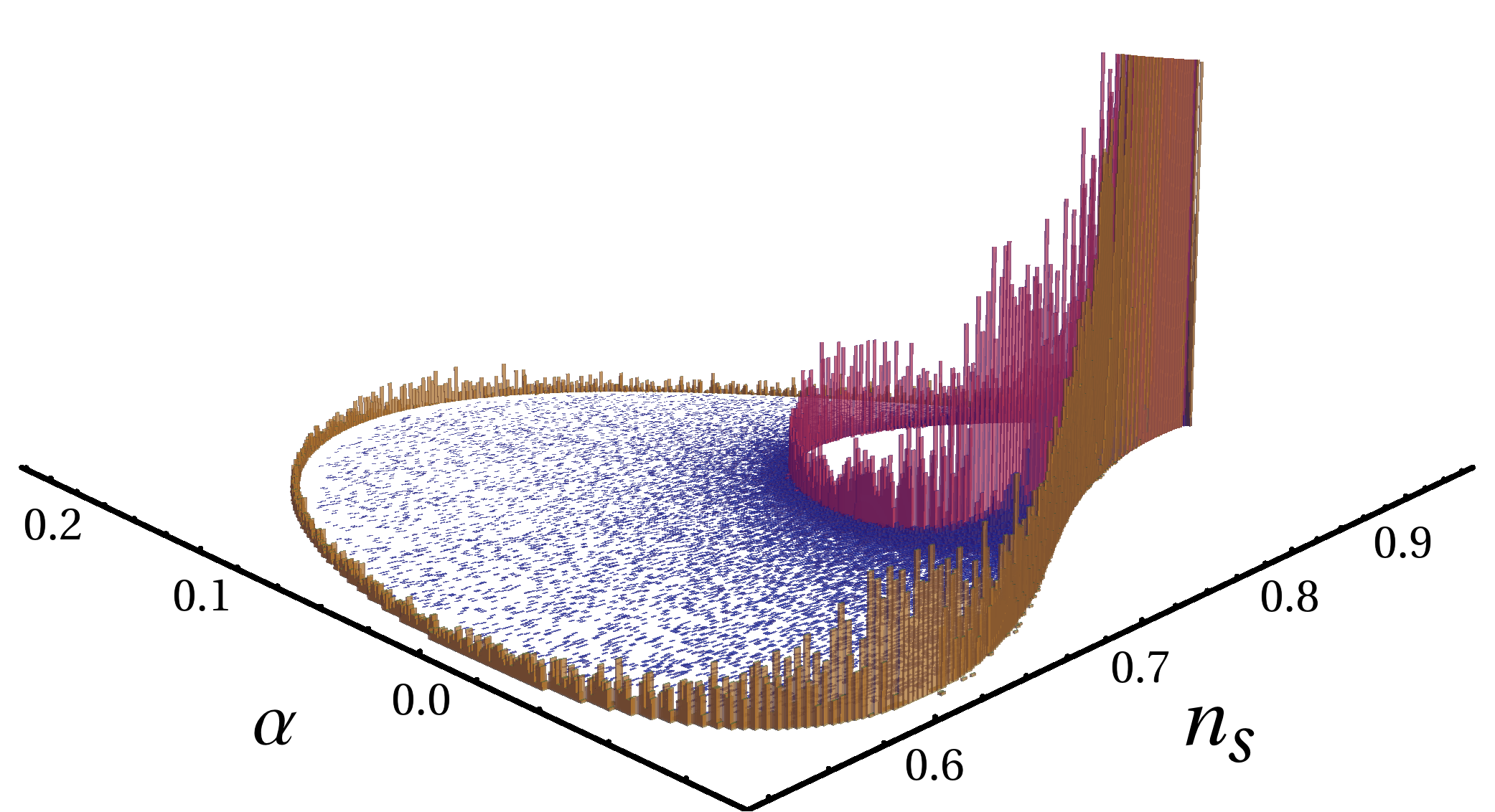}
  \caption{Histograms for the spectral index $n_s$ and running $\alpha$ for $N$-quadratic inflation with equal-energy initial conditions at $E_0=0.1 \, \mpl$. (Top) Three fields with mass ratios $m_i/m_1=\{1,7,9\}$; (Bottom) comparison between three fields (blue; masses as above) and two fields with mass ratio $m_2/m_1=7$ (red; inner contour) and $m_2/m_1=9$ (gold; outer contour).  The bottom figure emphasizes the outlying regions and does not show the full range.  All sampled distributions have a peak near $n_s=0.963$ and $\alpha=-7\e{-4}$, with appreciable deviation only in the tails. }
  \label{fig:nsalpha_3field_EE}
\end{figure}

\begin{figure}
  \includegraphics{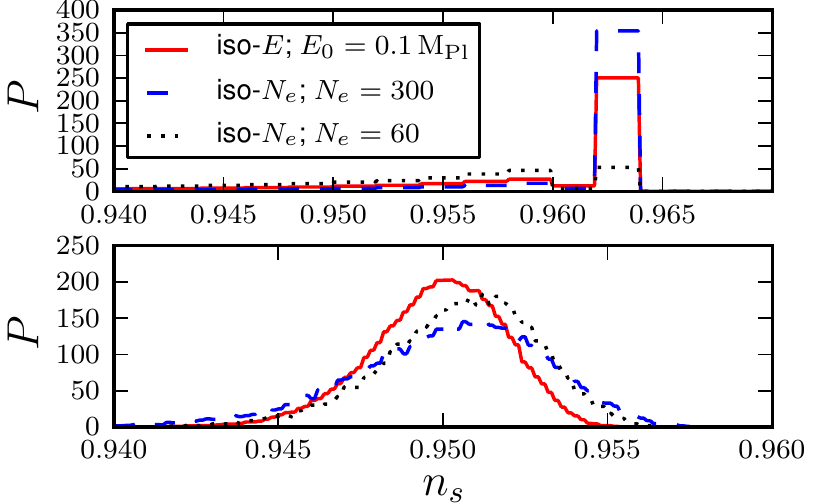}
  \caption{Probability distribution for (top) $N=3$ fields and (bottom) $N=100$ fields with different initial conditions (ICs) priors.  The slow-roll prior, which overlaps the iso-$E_0$ prior, has not been plotted. Importantly, the upper plot shows only the peak in the PDF and the full range for $n_s$ has not been plotted. The distributions show remarkable consistency, despite significantly different IC priors.}
  \label{fig:cdf_100field_EE}
\end{figure}

{\bf Results:} Figs.~\ref{fig:nsralpha_100field_EE}--\ref{fig:nsalpha_3field_EE} show histograms for $n_s$, $\alpha$, $r$, and $\riso$, with Scott-binning \cite{scott1979optimal} for the iso-$E_0$ prior with $E_0=0.1 \, \mpl$.   Fig.~\ref{fig:cdf_100field_EE} is the histogram-estimated PDF for $n_s$ for different IC priors.  The slow-roll prior yields results that are effectively indistinguishable from the iso-$E_0$ prior and are not plotted.  We sample $\mathcal O(10^6)$ ICs for $N=\{2,3\}$ fields and $\mathcal O(10^4)$ ICs for $N=100$.

Fig.~\ref{fig:nsralpha_100field_EE} shows the first-ever general predictions for $n_s$, $\alpha$, and $r$ for inflation with $N=\mathcal O(100)$ light fields.  Using the iso-$E_0$ prior, we find that $95\%$ of Monte Carlo samples are in the ranges: $n_s \in (0.9455,0.9534)$; $\alpha \in (-9.741,-7.047)\e{-4}$; $r\in(0.1445,0.1449)$; and $\riso \in (0.02137,3.510)\e{-3}$, which is similar to the predictions of this model in the single-field limit.  Crucially, while many-field $N$-quadratic inflation supports a  broader range of \emph{possible} observables, it nevertheless makes a sharp \emph{generic} prediction for $n_s$, $\alpha$, and $r$.

The $\riso$ component is significant with $N=100$ fields for the two- and three-field cases.  The sensitivity of $\riso$ on the choice of IC prior is shown in Fig.~\ref{fig:r_iso}.  The isocurvature fraction is largest for the iso-$N_e=60$ prior, reflecting the relatively short period this models has to evolve before inflation ends. The average number of total $e$-folds for the iso-$E_0=0.1\, \mpl$ prior is $N_e = 306.6$, and the average $\riso$ roughly decreases with increasing total number of $e$-folds.    Unlike the case of a few fields, the heavier fields do not always reach their minima before inflation ends, although they approach their minima given more time to evolve.  Trajectories are therefore typically turning in field-space at the end of inflation, and it is known \cite{Frazer:2011br,Seery:2012vj,Huston:2013kgl,Peterson:2010np} that this causes the isocurvature modes to grow. We attribute the increase in $\riso$ to these dynamical effects.

Fig.~\ref{fig:nsalpha_3field_EE} shows the PDFs for the observables for $N=3$ with an iso-$E_0$ prior with $E_0=0.1 \, \mpl$. The PDFs have  spikes in the bin $n_s \in(0.962,0.964)$ and $\alpha \in (-0.001,-0.0005)$, which contains $48.8\%$ of the Monte Carlo samples.  With $N=2$ we find contours in the ($n_s$,$\alpha$)--plane, reproducing the  analytic result of Ref.~\cite{Frazer:2013zoa}. For three fields the distribution is bounded by the same contours, with a lower weighting around the outer contour (with $m_2/m_1=9$). Typical trajectories become effectively single field before the end of inflation and hence isocurvature modes decay, giving negligible $\riso$.

To explicitly compare different IC priors, Fig.~\ref{fig:cdf_100field_EE} plots the prediction for $n_s$, the observable best-constrained by \emph{Planck}.  For many fields, the distributions are largely similar and are well-described near the maximum by Gaussians with means $\mu=(0.950,0.951,0.951)$ and variances $\sigma^2=(1.97,2.81,2.24)\e{-3}$, for the iso-$E_0$, iso-$N_e=300$, and iso-$N_e=60$ priors, respectively.  

With fewer fields, we see larger differences in the PDFs.  Nevertheless, the bin containing maximum probability mass coincides and all the PDFs have the same overall shape.  For the iso-$E_0$ and iso-$N_e=300$ priors the probability-mass lies in a small range of observable-space, giving essentially the same prediction.  Furthermore, as seen in Fig.~\ref{fig:nsalpha_3field_EE}, the outlying contours non-trivially constrain the joint prediction for $(n_s, \alpha)$.  For these two IC priors, the PDFs in Fig.~\ref{fig:cdf_100field_EE} do not change drastically.

However, the iso-$N_e=60$ prior (which is the least predictive choice \emph{a priori}) has a significantly lower peak in Fig.~\ref{fig:nsalpha_3field_EE} and 95\% of Monte Carlo samples in the broad range $n_s \in (0.675,0.963)$, which is comparable to the full range of predictions for this model, $n_s \in (0.543,0.964)$.  Interestingly,  this IC prior yields significant mass below the most probable value of $n_s=0.963$, and may thus perform far worse relative to the other IC priors in a Bayesian evidence calculation, as the 68\% \emph{Planck} bounds are $n_s \in (0.954,0.973)$.  This implies that \emph{Planck} data may constrain the  initial states when $N$ is small.

{\bf Discussion:}   This \emph{Letter} presents a complete analysis of multifield quadratic inflation. We numerically integrate the multifield mode equations through to the end of inflation, the first time this task has been performed for a model with many degrees of freedom. The code will be released and described separately \cite{MultiModeCode}. We compute PDFs for key observables, and evaluate their sensitivity to priors for the initial field values and velocities. 

We find that the initial conditions are not ``stiff parameters'' \cite{PhysRevLett.99.100602,Transtrum:2010zz,Machta01112013} for which small changes  radically alter observables,  demonstrating that this model makes sharp, robust predictions for the inflationary parameters.  Given that multifield models can produce a wide range of perturbation spectra, one may specify an IC prior for which the observables are far from the peak values in the PDF found here. However, such scenarios are typically  contrived, so the corresponding prior is unlikely to be physically compelling.  Moreover, even with one field, initial conditions which violate slow-roll near $N_e=60$ yield a $\PR$ that differs significantly from the usual result.

The matching between the number of $e$-folds and present-day scales depends on the post-inflationary equation of state \cite{Adshead:2008vn,Mortonson:2010er,Easther:2011yq}. This resulting uncertainty in $n_s$ and other observables scales with $\alpha$  and  is comparable to the width of the large-$N$ PDFs computed here. Consequently, the spread in the predictions of the inflationary observables at large $N$ --- including the ambiguity associated with the IC prior --- need not be the dominant source of theoretical uncertainty in  multifield models. 

For $N=100$   the isocurvature modes are potentially nontrivial. This is a new and significant result: the presence of isocurvature modes implies that the curvature perturbation may continue to evolve until an adiabatic limit is reached \cite{GarciaBellido:1995qq,Elliston:2011dr,Seery:2012vj,Frazer:2011tg,Dias:2012nf}, even if this is after the end of inflation.  However, the most probable values for the power spectra observables \emph{at the end of inflation} are still concentrated in small regions.  Recent studies of the evolution of observables during reheating focus on models with only a few fields  \cite{Leung:2012ve,Huston:2013kgl,Leung:2013rza,Meyers:2013gua}. Given the magnitude of the $\riso$ for $N=100$, it will be important to examine the reheating dependence of observables at large $N$, for which a non-zero $\riso$ may be generic.

With $N=100$, the central values we find for $n_s$, $r$ and $\alpha$ are consistent with those seen in previous work \cite{Easther:2005zr,Dimopoulos:2005ac,Kim:2006ys,Kim:2007bc} based on slow-roll expressions.  If the duration of inflation is increased by changing the initial conditions while other parameters are held fixed, $\riso$ is reduced, consistent with Fig.~\ref{fig:r_iso}.  However, there is no generic mechanism that forces the initial values of $\dot{\phi}_i$ to be  small \cite{Easther:2013bga} and, with the exception of our {\em slow roll prior\/}, we start our simulations with significant field velocities, in contrast to previous work. This reduces the  duration of inflation at fixed initial energy, and  increases the likelihood of seeing a nontrivial value of $\riso$.

Importantly, our results suggest that the curvature perturbation of multifield inflationary models has a well-defined large-$N$ limit. Consequently, these models may be  least predictive when $N=2$ or $3$. This situation mirrors that found elsewhere \cite{Aazami:2005jf,Easther:2005zr} and can be understood by analogy with the central limit theorem. Determining the extent to which this phenomenon is generic  in $N$-field inflation is clearly of the utmost importance. Finally, this analysis points the way to constraining multifield scenarios with data from observational surveys, such as {\em Planck}. 

\mbox{}

{\bf Note Added.---} After this Letter was completed a detection of primordial $B$-mode polarization in the CMB was announced by the BICEP2 collaboration  \cite{Ade:2014xna}.  The primary  goal of this Letter was to investigate the dynamics of a representative multifield model rather than to propose a candidate model of cosmological inflation.   However, we note that the model analyzed here predicts $r \sim 0.145$ and  permits a significant running, in qualitative agreement with  BICEP2. Moreover, different models of foreground dust subtraction further improve the fit, as illustrated in Fig.~11 of Ref.~\cite{Ade:2014xna}.

\mbox{}

\acknowledgments

\paragraph{Acknowledgments.---}
We thank Grigor Aslanyan, Mafalda Dias, Andrew Liddle, David Mulryne, and David Seery for useful discussions and Jiajun Xu for collaborating on the development of \textsc{ModeCode}.  JF is supported by the Leverhulme Trust, by IKERBASQUE, the Basque Foundation for Science, and by a grant from the Foundational Questions Institute (FQXi) Fund, a donor-advised fund of the Silicon Valley Community Foundation on the basis of proposal FQXiRFP3-1015 to the Foundational Questions Institute.  HVP is supported by STFC, the Leverhulme Trust, and the European Research Council under the European Community's Seventh Framework Programme (FP7/2007-2013) / ERC grant agreement no 306478-CosmicDawn.  We acknowledge the contribution of the NeSI high-performance computing facilities and the staff at the Centre for eResearch at the University of Auckland. New Zealand's national facilities are provided by the New Zealand eScience Infrastructure (NeSI) and funded jointly by NeSI's collaborator institutions and through the Ministry of Business, Innovation \& Employment's Research Infrastructure programme \footnote{{\url{http://www.nesi.org.nz}}}. This work has been facilitated by the Royal Society under their International Exchanges Scheme.

\bibliographystyle{apsrev4-1}
\bibliography{references}

\end{document}